# AMDA: Matching the Model-Driven-Architecture's goals using Extended Automata as a common model for design and execution

D. Dayan[1,2,3], R. Kaplinsky[4], A. Wiesen[2], S. Bloch[3]

[1]Sami Shamoon College of Engineering, Industrial Eng. and Manag. Dept, Bialik/Basel Str.,
Beer Sheva 84100, Israel
[2]Jerusalem College of Technology - POB 16031- Jerusalem 91160, Israel
[3]Univ. Reims, CReSTIC, Reims, France
[4]Jerusalem College of Engineering, POB 3566, Jerusalem 91035, Israel
david674@bezeqint.net, rkaplins@ort.org.il, wiesen@jct.ac.il,
simon.bloch@univ-reims.fr

## Abstract

*This paper proposes a model of execution platform for the OMG request of a generic Platform-Independent-Model (PIM) allowing realization of the Model Driven Architecture (MDA) standard.*

*We propose AMDA (Automata based MDA), a method based on the use of parallel automata, which can be a common tool for building a PIM from UML diagrams (including OCL) and transforming the PIM to PSM automata and further to compilable code. Each platform would then have a mechanism to execute the translated code.*

*Our architecture for a general PSM translator of these automata allows portable execution on various specific implementation platforms. This general translator must be written, once, for the languages and with the libraries of the required specific PSM. This also allows interoperability between different PSMs. An ATM case study example is presented to illustrate the approach.*

*Keywords: MDA, UML, Extended Automata, XSLT.*

## 1. Introduction

OMG (Object Management Group) is an organization which proposes standards to unify the specification and the design of applications. For that, it developed the UML standard.

The Unified Modeling Language (UML) [1] adopts a pluralistic attitude toward the multiplicity of notations. Several diagrams and notations are incorporated within UML, addressing various aspects of system development. UML is not a method for executing the models, and thus does not address the way these diagrams are to be used. On the contrary, there are several diagrams, each one with its own distinct syntax and semantics, which can be used interchangeably to convey different views of the same information (e.g., Statecharts and Activity Diagrams, Sequence Diagrams and Collaboration Diagrams).

OMG proposed a new standard: MDA (Model Driven Architecture) which aims to separate application logic from underlying platform technology, so that the applications are platform independent and can be realized on various underlying platforms (including J2EE, .NET, Web-based platforms etc.). So platform-independent application models can help to free the application development from technology specifics and allow easier interoperability between various applications and platforms.

Let us quote the OMG objectives [2]: "*The MDA is a new way of writing specifications and developing applications, based on a platform-independent model (PIM). A complete MDA specification consists of a definitive platform-independent base UML™ model, plus one or more platform-specific models (PSM) and interface definition sets, each describing how the base model is implemented on a different middleware platform. A complete MDA application consists of a definitive PIM, plus one or more PSMs and complete implementations, one on each platform that the application developer decides to support.*"

Jon Siegel, Director of Technology Transfer at OMG says [3]: "*For platform independence, OMG will standardize – and MDA tools will implement – mappings to multiple middleware platforms. Each mapping – formally, a UML profile – defines the route from an application's single PIM to a PSM on a target platform, i.e. UML profiles will be mapped through a PIM to middleware technology*". This view is summed up in Figure 1 (quoted from [3]).

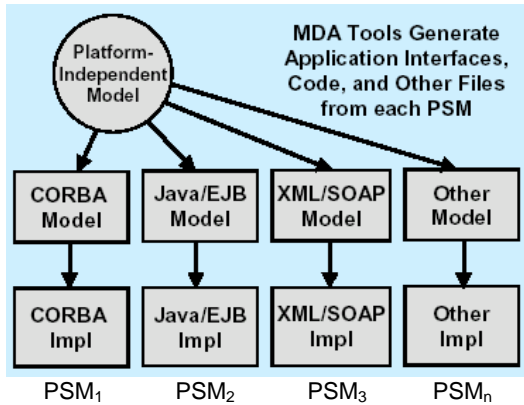

**Figure 1. MDA portability through PIM-PSM**

In this view, the middleware technology is at the level of the implementation system (PSM for CORBA, Java, XML etc.).

## 2. AMDA: Our PIM/PSM Software Proposal

With AMDA, we propose a particular approach to the generic PIM-PSM transition problem: the Extended Automata model. The modeled application PIM is based on these automata and is a result of an automatic translation of the applicative UML diagrams, persisting in XML format.

In order to facilitate transition from automata-based PIM to automata-based PSM, a middleware translator is required. AMDA addresses this task using XSLT transformation.

The final transformation to compilable source code is performed in AMDA with the aid of XSLT and additional tuning instructions of the tool.

So, we think that it is possible that the PIM will be a small software layer between the UML application Model and the implementation systems (Java, .NET, XML etc.). This PIM software layer will have two objectives: a) it will be an intermediate translation of the UML Model of the application, similar for instance to the "*byte code*" which is an intermediate translation of Java source text, and b) it will be interpreted by the various implementation systems.

This software PIM will be a real interpretation middleware between the application model and the implementation systems. The applicative model translation will be unique, and the execution will be specific for each implementation system (each one will have its adapted interpreter), see Figure 2.

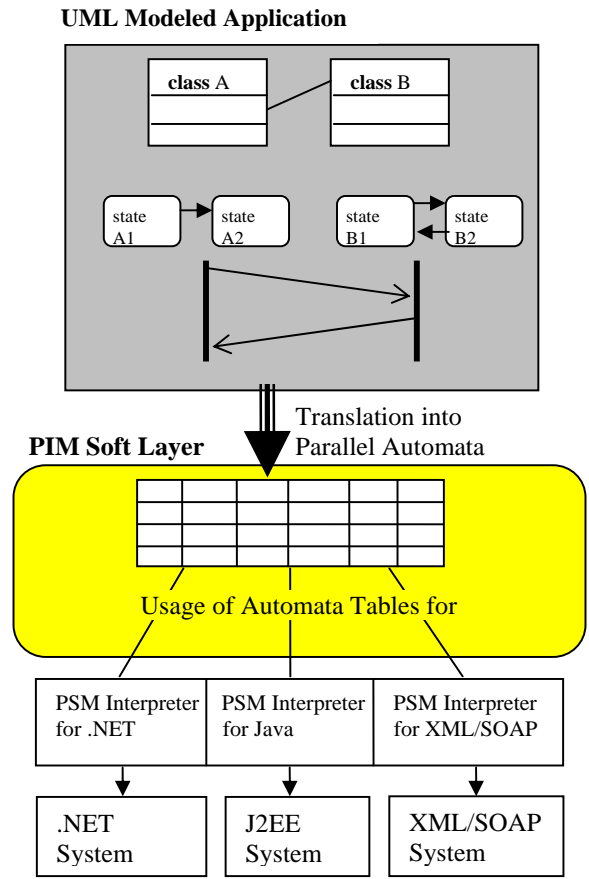

**Figure 2. PIM software intermediate layer**

Furthermore, the coordination and the synchronization between various systems will be simpler. It will be made at the level of the applicative models, i.e. at the level of the translated PIM intermediate layer.

The dynamic behavior of the system objects is captured by the UML statechart diagrams, which are based on the notion of statecharts, introduced by David Harel [4]. Many different variants of Harel statecharts are known from the literature [5], and various formalizations were proposed, such as Extended Hierarchical Automata (EHA) [6] and the Parallel Automata [7].

We have based our approach to MDA's transformations on building PIM out of blocks which correspond to objects modeled by UML state diagrams. Each such block is an extended automaton of a kind we introduce in this section, namely, Statechart Sequential Automaton (SSA), Hierarchical Sequential Automaton (HSA) and Parallel Hierarchical Sequential Automaton (PHSA). Each of these models is an expansion of the previous.

## 2.1 The Statechart Sequential Automaton

The SSA includes four components (see Fig.3): the first is a reactive component $\mathscr{A}$ which is essentially a Moore automaton, whose states correspond to those of the UML statechart and the role of input alphabet play combinations of an incoming event and a guard (Boolean expression involving system variables). The output symbols are actions performed by the automaton when it enters a destination state. Second component is a stateless transformational scheme $\mathscr{C}$, which centralizes computation of conditions and is responsible for executing local methods. The third is a memory register $\mathscr{M}$ for storing the system variables. The fourth component, I/O, performs input-output routines.

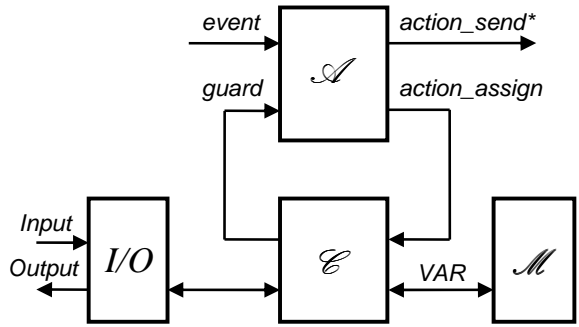

**Figure 3. The SSA structural scheme**

This decomposition helps us to separate the platform-independent part of the modeled system (the Moore automaton) from other components, which are to some extent platform-dependant: implementation of the memory and the conditional scheme depends on data types supported in the target platform, and the I/O system is completely platform-dependant.

## 2.2 Extending the SSA model to include hierarchy (composite states)

The HSA is an SSA extension that treats composite states of a UML statechart, and PHSA goes forth to deal with parallel execution of the sub-automata.

An application defined by UML statechart may contain composite states. Such states are themselves statecharts, so the application can be represented by a hierarchical structure. In general, a composite state may contain several "inner" statecharts, which in this case must be executed concurrently, but at this step of the modeling we suppose that each composite state contains exactly one statechart, that is, we deal at this step only with hierarchy, but not with concurrency (parallelism). Here, we describe a formal model that we call Hierarchical Statechart Automaton (HSA).

Let us consider a hierarchy of $K+1$ automata made of a main automaton $A_0$ and sub-automata $A_1$, ..., $A_K$. The only restriction we put on the structure of the system is that it has to be a tree, that is, each sub-automaton belongs only to one parent.

The hierarchical behavior of a system means that each transition is on its definite level, so each transition changes the state of *one* definite sub-automaton only, and access to other sub-automata allowed only through their initial states. Therefore, each transition takes the form:

$$state_{km} \xrightarrow{event_i \, / \, cond_j} state_{kn}, k = 0,1,...,K , \quad (1)$$

where $state_{km}$, $state_{kn} \in STATES_k$ (the set of the states of the automaton $A_k$).

We assume that the statecharts $CHART_0$, $CHART_1$, ... , $CHART_K$ of the main automaton $A_0$ and its sub-automata $A_1$, ..., $A_K$ are given. Some of the vertices of $CHART_0$ (i.e. states of $A_0$) are composite states of $A_0$ and are interpreted as sub-automata from the list $A_1$, ..., $A_K$; the same may occur for some of the states of any one of sub-automata. Our restriction means that all the connections in the system are arranged in a tree.

For each statechart $CHART_k$ there is a subset $H_k$ of its vertices that present composite states of $A_k$ that are to be interpreted as sub-automata from the same list above $A_1$, ..., $A_K$, i.e. we mean that there is given the set of functions $f_k$:

$$f_k : H_k \rightarrow \{1, ... , K\} \quad (2)$$
$$state_{k\alpha} \mapsto f_k(\alpha)$$

Each function $f_k$ maps the indices of the composite state in $A_k$ into the indices of their corresponding automata on the next (lower) level of hierarchy. In the simplest case, we suggest that no automaton in the list may correspond to more then one composite state of any other automata, i.e. our net of automata form a tree structure.

Now, in order to represent hierarchical automaton (HSA) as composition of components which are SSA automata, we add for each composite state $state_{k\alpha}$ four new elements:

1. entry action $DummyAction_{k\alpha}$ to start execution of the sub-automaton $A_{f_k(\alpha)}$

2. "dummy" state $DummyState_{k\alpha}$

3. event $DummyEvent_{k\alpha}$ that each sub-automaton $A_{f_k(\alpha)}$ produces when it reaches its final state

4. transition from the state $state_{k\alpha}$ to the $DummyState_{k\alpha}$

The purpose of the $DummyAction_{ka}$ is to move down to the lower level of hierarchy, while the rest three two elements cause the automaton to return to the previous level.

Introducing the dummy states is simply made by extending the original set of UML state diagram states $S_k$ with:

$$\widetilde{S}_k = S_k \cup (\bigcup_{i \in H_k} \{DummyState_{ki}\}) , \qquad (3)$$

That is, for every automaton $A_k$ which contains a set $H_k$ of composite states, the set $S_k$ has to be extended with only one dummy state for each composite state.

Appropriately, we will need to add new entities in the XML document representing the table of events in the serialized form of our automaton.

$$\widetilde{E}_k = E_k \cup (\bigcup_{i \in H_k} \{DummyEvent_{ki}\}) \qquad (4)$$

The event $DummyEvent_{ki}$ must be triggered when the sub-automaton $A_{f(k,i)}$ reaches its final state. This requires adding an entry action of "send event" kind in the final state of every sub-automaton in all the hierarchy. In this way we can reduce each composite state in UML statechart to a composition of SSA blocks.

## 3. AMDA Work Outline

After we have described our formal model of extended automata and the representation of UML statecharts as extended automata, let us determine the role of the automata in MDA-oriented development process, as shown in Fig. 4.

### 3.1 Transformation of a UML diagram's XMI File to an Automata-based PIM

In our vision of this process we have been guided by various sources related to OMG, of which is worth to mention [8] and [9]. In order to allow open development process, we have to export these diagrams to XMI format, which is standard de facto for interchanging XML and UML documents and is supported by various tools, e. g. I-Logix Rhapsody, Rational Rose, IBM WebSphere, Borland Together and others. The UML models and exported XMI files contain OCL constraints and expressions [10], [11].

At the first stage of application modeling, the designer builds UML diagrams. In our tool AMDA, we use three kinds of diagrams: class, state and sequence. These diagrams capture both structural and dynamic aspects of the modeled application.

The UML diagrams are exported to XMI documents, which are the input to AMDA. The tool reads XMI, strips all irrelevant information such as geometry and colors, and creates XML documents according to our PHSA automata model. The contents of the PHSA components for each object are in the tables we present in the use case.

According to the PHSA automata model, the platform-dependent and platform-invariant components are separated to facilitate transformations to PSMs. Further, it is possible a) to simulate the behavior of the modeled application in order to check its functionality, b) to export PHSA tables in XML format according to DTD we have defined, and c) to transform the PIM to PSMs using XSLT style sheets for various platforms (.Net and J2EE).

## 4. Automata PIM Structure and Execution Semantics

Now we will describe the use of our formal automata model to define the PIM structure and semantics. We create our PIMs as XML documents according to DTD that reflects PHSA structure. Since in PHSA the platform dependant and platform independent parts are already separated, this technique facilitates further transformations to PSMs.

### 4.1 PHSA Automata Realization in XML

The PHSA building blocks are SSAs which, as described above, consist of four main components: the Moore automaton, the condition scheme, the memory and the input-output system.

a) the Moore automaton is defined in the DTD in `automat` element. In this part we write all the states belonging to the automaton, the events that this automaton receives from other objects, and all the transitions of the automaton. The events that the automaton sends will be written as entry actions.

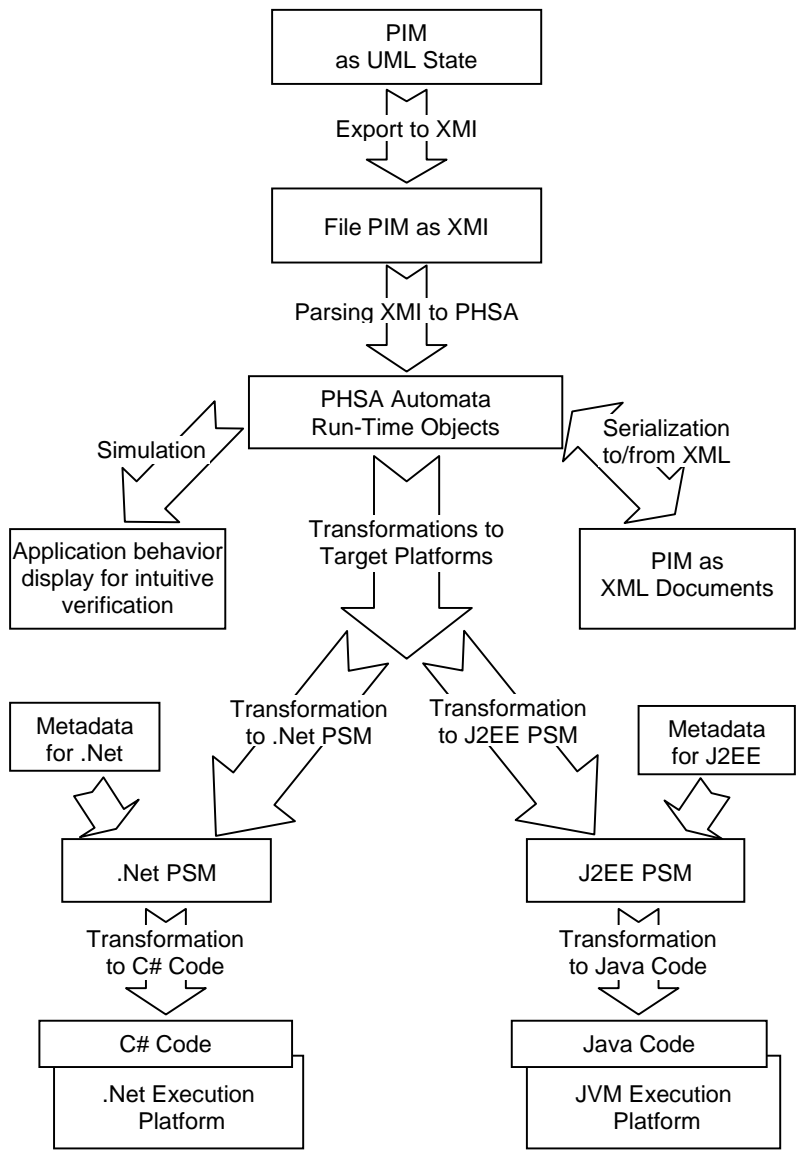

**Figure 4. Integration of PHSA extended automata in MDA process**

Each automaton includes at least two states: the initial and the final pseudo-states. Each state has identifier and name. On entering a state, the automaton may produce some entry actions, which are output symbols in terms of Moore automaton. An entry action may be of one of the three kinds: *inline*, *function* or *send-event*. An *inline* action, defined by `act_inline` element, is a simple instruction like an assignment; its body is written immediately within the `state` element. A *function* action, defined by `act_func` element, is essentially a function, which is called through a function identifier. The `state` element contains only the id, and the function body appears in the condition scheme (see below). A *send-event* action, defined by `act_send_event` element, is the action that permits sending events to other objects.

b) the condition scheme is defined in `condscheme` element. This part is responsible for evaluating expressions (guards) and performing computational actions, i.e. of *function* kind. It consists of two collections: conditions and `func_actions`. Each condition element has identifier and body. The id references the condition from a transition where it occurs as a guard, and the body contains a boolean OCL expression. The `func_action` element has a unique id to be referenced from a state as an entry action, and a function body.

c) the memory is defined in `memory` element. This part stores the PHSA variables (the object's data members). A variable is defined by its name, type and initial value. Types supported for now are `integer`, `real`, `flag`, `char`, `string`, `ord_collect` and `unord_collect` (for ordered and unordered collections). The initial value is an OCL expression.

d) the input-output system is defined in `iosystem` element. This is a virtual driver for performing input-output operations. An input is read into a variable stored in the memory, while an output can be any expression. We implemented two modes of input-output: stream and GUI. The input-output operations are regarded as entry actions and are executed from the condition scheme. The `iosystem` element contains a collection `io_actions`, each member of which can be either `i_action` or `o_action`. In both cases it has following attributes: the operation id (to be referenced from the condition scheme), the mode of operation, reference to the variable or expression and the destination of input or output.

## 4.2 Application Execution Sequence in XML

In addition to defining the structure and behavior of single objects captured by PHSA automata, we have to define objects' instantiation and their interaction, i.e. methods calls. This information is supplied in additional XML file, which reflects the dynamic aspect of the application and is generated from the sequence diagram represented as XMI file. We call this file the "application dispatcher" file.

## 5. Transformation from PIM to PSM

At this stage we have the PIM of the whole application in form of in XML files containing PHSA definitions and the "application dispatcher". We will explain the rules of transformation from the PIM to PSM.

To define the transformation rules we have chosen to use XSLT [12] in combination with Octopus OCL processor [13]. Thus, transformation definition for each specific platform will be in form of an XSL stylesheet.

### 5.1 Transformation of PHSA Components

Since the reactive component of PHSA (the Moore automaton) is platform-independent, it is copied as is to the destination PSM XML file.

The memory register is but slightly dependant on the platform. In order to transform it, we have to specify concrete data types and structures that support generic types (listed in 4.1, item c) in the list). For example, the generic type `flag` in PIM will be translated for Java platform as boolean, and for .Net platform as `bool`. The generic type `ord_collect` will be translated to `ArrayList` in Java and to `Array` in .Net. The initializers are OCL expressions and are interpreted by the OCL processor.

The condition scheme contains definitions of conditions (guards) and function bodies. Both are in OCL, with the difference that the first are Boolean expressions and the second are routines.

The input-output system is totally platform-dependant PHSA component. In case of stream input-output, we specify the classes and methods that support appropriate stream types and input-output operations in the target platform. For example, console output in Java will sound as `System.out.print(myVar)` and in .Net as `Console.Write(myVar)`, while the console input in Java is a bit more complicated and requires more than a single command, so we have encapsulated it in a small helper class. Thanks to the unified stream input-output mechanism, destination stream can be not just console, but also a file, a socket etc. For GUI input-output, each variable is associated with an appropriate control, e.g. a textbox for a string and a checkbox for a flag.

### 5.2 The Transformation Rules (PIM to PSM)

The transformation rules are:
1. Write to the PSM import statements for all the needed framework packages.
2. Write import statements for generic collection classes.
3. The `automat` element is copied to PSM unchanged.
4. The `condscheme` consists of two parts: `conditions` and `actions`.
4.1 The conditions are OCL expressions and are copied unchanged, since OCL is translated directly to code during the transformation from PSM to code.
4.2 Actions : The algorithms of the actions must be already written before the translation.

5. Translation of the `memory` element: for each `variable` element the `type` is translated to appropriate platform-specific basic type or data structure.

6. The input-output component, `iosystem`, consists of input and output actions.

6.1 For each input action:

6.1.1 If the mode is text, write the PSM input statement, specifying the stream class and its method responsible for text input on the target platform, and the variable to input.

6.1.2 If the mode is GUI, the variable value is read from an input dialog. The PSM input statement is similar to 6.1.1, but the platform-specific input dialog class and its appropriate method are used instead of the stream class.

6.2 The text and GUI output actions are similar to input actions (as in 6.1) with the difference that parameters in output statements are expressions and not just variables.

## 5.4 Transformation of Associations

Associations between objects are captured by UML class diagrams. OCL supports association roles, navigation and multiplicities. The OCL constraints and queries can be readily translated to Java or to .Net platforms.

## 5.5 Single File vs. Split Transformation Definitions

There are two possibilities: to write all the transformation rules in a single file, or to split them into several files, or "libraries" of transformation rules. For example, it is possible to write a library of basic functions, input-output operations, and libraries of application-specific functions. Both methods have their pros and cons we don't have place to discuss here. Our choice is a single-file transformation definition, but it allows attaching additional libraries.

## 6. Transformation from PSM to Code and Execution

After we get PSM targeted at a specific platform, transformation to code is rather straightforward. We have to create files containing class definitions for all the PHSA components and to instantiate PHSA objects.

In more detailed view, following classes are to be created:

**Table 1. PHSA Classes**

| PHSA Components | Aggregated Objects |
|---|---|
| Main PHSA | State, Event, Guard, Transition, Action |
| Memory | Variable |
| Condition scheme | FunctionalAction, OclExpression, SendEvent |
| I/O System | Window, Console, StreamInput, StreamOutput, GuiInput, GuiOutput |

## 6.1 The PSM to Code Transformation Rules

For each PSM PHSA automaton is generated class definition in a separate source file. All specific PHSA classes inherit the abstract PHSA class which implements the generic PHSA structure. The "application dispatcher" class is generated from its XML file (see 4.2).

The transformation rules are:

1. The import tags are translated into Java import statements.

2. The OrderedCollection and UnorderedCollection tags in the FoundationClasses element define the types used for PHSA inner components and variables that are collections. Examples of unordered collections are states and transitions, while actions must be an ordered collection, since order of execution matters. Each occurrence of OrderedCollection or UnorderedCollection in this sample is translated to ArrayList or HashTable respectively.

3. Each PHSA element is translated to its class definition according to definitions in XML file. All PHSA classes are derived from the abstract ClassPHSA. Within these classes are generated following data members and methods:

3.1 From each automat XML element is generated a class, which implements the Moore automaton behavior. The name of the class is defined by appropriate attribute of the element. The class contains data members states and transitions, both are unordered collections. The transition function is implemented through the handler method (see code snippet in section 6.7).

3.2 From the condscheme element is generated the ConditionScheme class. It contains unordered collections of Guard objects and FuncAction objects.

3.3 From the memory element is generated the Memory class. It contains an unordered collection of Variable objects generated from variable elements.

3.4 From the iosystem element is generated the IOSystem class. It contains an unordered collection of IOAction objects. Each action object may be of one of four types: input or output and stream or GUI.

## 6.2 Execution

The generated program is in fact a working simulation of the modeled system. In order to use it in a real system, there must be provided some program interface for interchanging events with the ambient environment on the target platform. To use the program as a simulation, e.g. for an intuitive visual verification, we added a GUI window offering to the user a list of possible external events with explanative descriptions (taken from appropriate attribute of the event XML element). Also are shown the current state, performed transitions and system variables. Input-output is performed from GUI controls, which in a real environment would be substituted with a real program interface using the same virtual driver.

## 7. Case Study: Automatic Teller Machine

We have prepared an illustrative case study on ATM (Automatic Teller Machine) which includes several use cases. In Figure 5 is given UML state diagram describing behavior of the ATM main logic controller in use case of client identification. The use case logic works as described below.

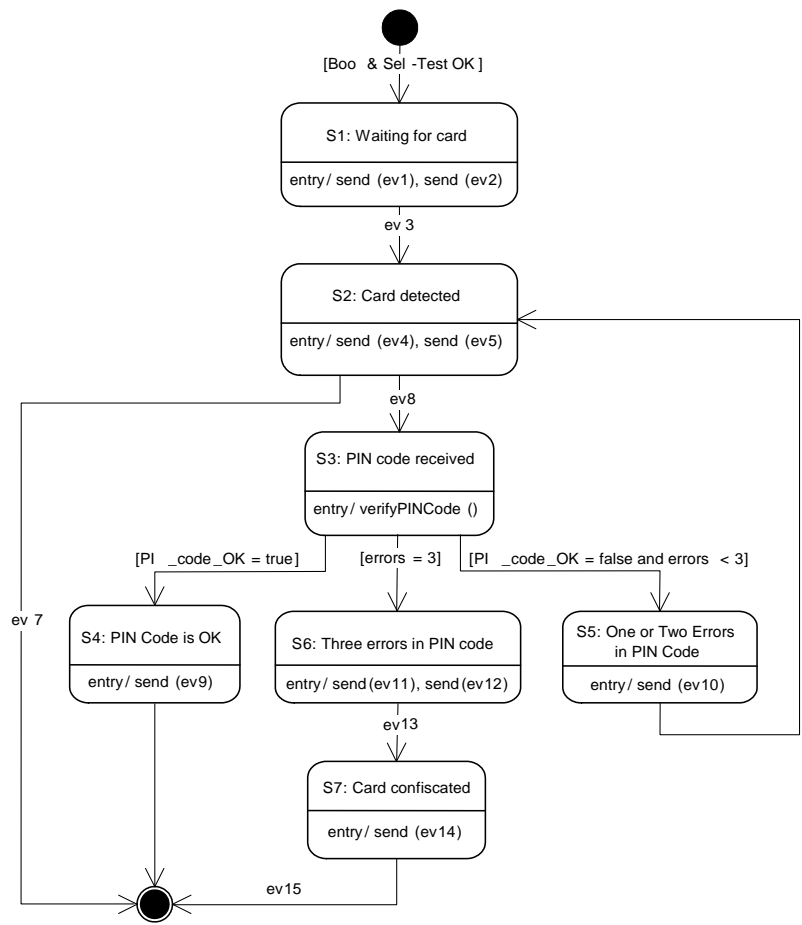

**Figure 5. UML state diagram of the ATM main logic controller in use case of client identification**

## 7.1 Application Logic

On completion the boot and self-test, the ATM waits in state `S1` until the user inserts a credit card. On entering `S1` it sends signal `ev1` telling the monitor to display the welcome message, and signal `ev2` to the card reader unit to start checking for a card. On receiving signal `ev3` from the card reader indicating that a card was inserted, it changes to the state `S2` and waits for the PIN code from the keyboard. On entering this state, the controller sends two signals: `ev4` to the monitor telling the prompt for the user, and `ev5` to the keyboard to start reading the PIN code. On receiving `ev8` from the keyboard telling that the all the digits of PIN code were entered, the controllers moves to `S3` and calls the routine `verifyPINCode` that checks if the entered code matches the code written on the credit card. If the code is OK, the signal `ev9` is sent to the monitor to display the user menu, and the use case of client identification is completed. If the enterd code is wrong, the user is given two more trials (`ev10` tells the monitor to display the "wrong code" message), and in case wrong codes were entered three consecutive times, the card is confiscated.

## 7.2 Synthesizing an SSA from the UML Statechart

From the UML state diagram of the ATM Main Controller object (Fig. 5) we get the following SSA components:

1. States table S={S0, S1, S2, S3, S4, S5, S6, S7, End}. The names S0 and End will be used constantly for the SSA initial and final pseudo-states.
2. Events table E= {ev3, ev7, ev8, ev13, ev15}.
3. Variables table V ={ errors, PIN_code_OK }.
4. Conditions table C = {
   [PIN_code_OK=true],
   [errors=3],
   [PIN_code_OK=false and errors<3]   }.
5. Assignment actions table AS = {
   v2 = true,
   v1= v1 + 1,
   verifyPINCode()  }.
6. "Send event" actions table
   SE = { send(ev1), send(ev2), send(ev4),
   send(ev5), send(ev9), send(ev10),
   send(ev11), send(ev12), send(ev14)  }.
7. Transitions table T: see Table 2.
8. Output table G: see Table 3.

### Table 2. Transitions table *T*

| Source state | Event detect | Test Condition | Dest. state |
|---|---|---|---|
| S0 | | | S1 |
| S1 | ev3 | | S2 |
| S2 | ev7 | | End |
| S2 | ev8 | | S3 |
| S3 | | [PIN_code_OK = true] | S4 |
| S3 | | [errors=3] | S6 |
| S3 | | [PIN_code_OK = false and errors < 3] | S5 |
| S4 | | | End |
| S5 | | | S2 |
| S6 | ev13 | | S7 |
| S7 | ev15 | | End |

### Table 3. Output table *G*

| State | "assign" action | "send event" action(s) (names of the events sent) |
|---|---|---|
| S1 | | ev1, ev2 |
| S2 | | ev4, ev5 |
| S3 | verifyPINCode() | |
| S4 | | ev9 |
| S5 | | ev10 |
| S6 | | ev11, ev12 |
| S7 | | ev14 |

## 7.3 Execution of the Tables on the Target Platform

The execution is performed in two stages, as described above. On the first stage we transform the PIM PHSA to PSM PHSA automaton using platform definition in form of XSL file. On the second stage we use analogous technique to end up with the compilable code. We present here fragments of the resulting XML files and code snippets for Java platform.

**PHSA PIM Automata in XML Format:**

```
<?xml version="1.0" encoding="UTF-8"
standalone="no"?>
<!DOCTYPE pim SYSTEM "pim_phsa.dtd">
<pim>
  <phsa phsa_id="A1">
    <automat>
      <states>
        <state phsa_ref="A1" state_id="A1_S0"
            state_name="S0">
        </state>
```

```
<state phsa_ref="A1" state_id="A1_S1"
        state_name="S1">
  <entry_action>
    <act_send_event event_id="ev1" />
  </entry_action>
  <entry_action>
    <act_send_event event_id="ev2" />
  </entry_action>
</state>
[…skipped…]
<state phsa_ref="A1" state_id="A1_S3"
        state_name="S3">
  <entry_action>
    <act_func act_id="A1_Func1" />
  </entry_action>
</state>
[…skipped…]
</states>
<events>
  <event event_id="ev3" />
  <event event_id="ev7" />
  <event event_id="ev8" />
  <event event_id="ev13" />
  <event event_id="ev15" />
</events>
<transitions>
  <transition state_src="A1_S0"
            state_dest="A1_S1"
  <transition state_src="A1_S1"
            state_dest="A1_S2"
            event_ref="ev3" />
  […skipped…]
</transitions>
</automat>
[ condscheme, memory and iosystem skipped ]
</phsa>
</pim>
```

## 7.4 Excerpt of PIM to PSM Transformation Definition for Java Platform (XSL Snippet)

```
<?xml version="1.0" ?>
<xsl:stylesheet version="1.0"
xmlns:xsl="http://www.w3.org/1999/XSL/Transform">
<xsl:template match="/pim">
  <psm_j2ee>
    <Imports>
      <import>java.io.*</import>
      <import>javax.swing.*</import>
      <import>amda.streamio.Console</import>
      […skipped…]
    </Imports>
    <FoundationClasses>
      <OrderedCollection>
          ArrayList
      </OrderedCollection>
      <UnorderedCollection>
          HashTable
      </UnorderedCollection>
      […skipped…]
```

```
    </FoundationClasses>
    <xsl:apply-templates select="phsa"/>
  </psm_j2ee>
</xsl:template>
[…skipped…]
<xsl:template match="variables">
  <xsl:for-each select="variable">
    <variable>
      <xsl:attribute name="psm_var_name">
        <xsl:value-of select="@name"/>
      </xsl:attribute>
      <xsl:attribute name="psm_var_type">
        <xsl:if test="@type='integer'">
          <xsl:text>int</xsl:text>
        </xsl:if>
        <xsl:if test="@type='flag' ">
          <xsl:text>boolean</xsl:text>
        </xsl:if>
      </xsl:attribute>
      […skipped…]
    </variable>
  </xsl:for-each>
</xsl:template>
[…skipped…]
</xsl:stylesheet>
```

## 7.5 PSM Automata for Java Platform (XML Snippet)

```
<?xml version="1.0" encoding="UTF-8" ?>
<psm_j2ee>
  <Imports>
      … same as in the XSL file (see 6.4)
  </Imports>
  <FoundationClasses>
      … same as in the XSL file (see 6.4)
  </FoundationClasses>
  <automat>
      … the automaton is copied from the PIM XML file
        (see 6.2)
  </automat>
  <condscheme> … </condscheme>
  <memory>
    <variables>
      <variable psm_var_name="errors"
          psm_var_type="int" init="0"/>
      <variable psm_var_name="PIN_code_OK"
          psm_var_type="boolean"/>
    </variables>
  </memory>
  <iosystem> … </iosystem>
</psm_j2ee>
```

## 7.6 Excerpt of Generated Code for Java Platform

```
public class PhsaA1 extends ClassPHSA  {
  private boolean PIN_code_OK;
  private int errors=0;
```

```
public void handler() {
    if (_cstate.equals("a1_s0")) {
            _cstate="a1_s1";
            a1_s1();
    }
    else if (_cstate.equals("a1_s1")) {
            if (_event.equals("ev3")) {
                _cstate="a1_s2";
                a1_s2();
            }
    }
    else if (_cstate.equals("a1_s2")) {
            if (_event.equals("ev7")) {
                _cstate="a1_end";
            }
            else if (_event.equals("ev8")) {
                _cstate="a1_s3";
                a1_s3();
            }
    }
    else if (_cstate.equals("a1_s3")) {
            if (PIN_code_OK==true) {
                _cstate="a1_s4";
                a1_s4();
            }
    }
}
[... skipped …]
```

## 8. Conclusion

The AMDA theoretical and technical approaches that are developed in this paper facilitate MDA process using UML state diagrams as an input and executable automata as output. On every step of our technique, the process is efficient in sense that we preserve the states of a source statechart and the only added states are minimal, so the model is transparent for an engineer and developer-friendly. Though the conflict analysis is beyond the scope of this paper, we can point out our model does not add any new conflicts compared to the source state diagram. The decomposition makes possible to use the notion of classical automaton with output (Moore machine) that we extend with two additional simple and well-known components, namely a stateless transformational scheme and a memory register. We also present a technique for handling concurrency. As next step of the work we intend to elaborate translation schemes (based on extended automata modeling) for the other design diagrams of UML and various execution platforms for achieving the MDA goals.